\documentclass[twocolumn,showpacs,preprintnumbers,amsmath,amssymb,superscriptaddress,floatfix]{revtex4}

\usepackage{graphicx}
\usepackage{dcolumn}
\usepackage{bm}
\usepackage{latexsym}
\newcommand{\beqn}{\begin{eqnarray}}
\newcommand{\eeqn}{\end{eqnarray}}

\newcommand{\Kanazawa}{\affiliation{Institute for Theoretical Physics,
Kanazawa University, Kanazawa 920-1192, Japan}}
\newcommand{\RIKEN}{\affiliation{RIKEN, Radiation Laboratory, Wako 351-0158, Japan}}

\begin{document}


\title{Gauge invariance of the dual Meissner effect in  QCD}
\author{Tsuneo Suzuki}
\Kanazawa
\RIKEN
\author{Katsuya Ishiguro}
\Kanazawa
\RIKEN
\author{Yoshifumi Nakamura} 
\Kanazawa
\RIKEN
\author{Toru Sekido}
\Kanazawa
\RIKEN

\date{\today}

\begin{abstract} 
The  dual Meissner effect is described and numerically observed in a gauge-invariant way in lattice Monte-Carlo simulations of pure $SU(2)$ QCD.
 A gauge-invariant Abelian-like field strength is defined in terms of a unit-vector in color space which is constructed by a non-Abelian field strength itself.   A gauge-invariant  monopole-like quantity  is defined by a violation of the  Bianchi identity with respect to the Abelian-like field strength. The squeezing  of the non-Abelian electric field $\sqrt{\sum_a(E^a_i)^2}$ between a pair of static quark and anti-quark occurs due to the solenoidal current coming from the gauge-invariant  monopole-like quantity. An equation similar to the dual London equation is confirmed approximately in the long-range region.    
 \end{abstract}


\maketitle

One of the most essential problems of color confinement in QCD is to explain the mechanism of the flux squeezing of 
non-Abelian electric fields between a pair of static quark and anti-quark.  In $SU(2)$ QCD, $\sum_a (E^a_i)^2$ or $\sqrt{\sum_a(E^a_i)^2}$ is expected to be squeezed to reproduce the linear static potential. Numerically the expected squeezing of the gauge-invariant combination of the electric field was observed beautifully in  lattice 
$SU(2)$ QCD~\cite{Bali:1994de}.

Thirty years ago, 'tHooft~\cite{tHooft:1975pu} and Mandelstam~\cite{Mandelstam:1974pi} conjectured that the dual Meissner effect is  the color confinement mechanism of QCD. 
However what causes the dual Meissner effect and how to treat the non-Abelian property were not clarified.

An interesting idea is to utilize a topological monopole like the 'tHooft-Polyakov monopole~\cite{'tHooft:1974qc, Polyakov:1974ek}.
An important quantity is a 'tHooft  field strength 
$\tilde{f}_{\mu\nu}=\sum_a n^aF^a_{\mu\nu}+\sum_{abc}\epsilon_{abc}n^a(D_{\mu}n)^b(D_{\nu}n)^c$ where $n^a$ is a unit vector  
composed of gluonic fields
transforming as an adjoint representation
in color space
and $(D_{\mu}n)^b$ is a covariant derivative.  
A monopole picture can be seen more clearly if we 
project 
$SU(3)$ QCD to an Abelian $U(1)^2$ theory by a partial gauge fixing~\cite{tHooft:1981ht}. 
Then we have an Abelian $U(1)^2$ theory with Abelian electric and magnetic charges. It is conjectured in Ref.\cite{tHooft:1981ht} that the condensation of the Abelian monopoles causes the dual Meissner effect explaining the color confinement. 

However there is a serious problem in this scenario.
Namely there exist infinite ways of choosing $n^a$ or in other words infinite possible Abelian projections. Moreover, the monopole condensation, if happens, can explain only the squeezing of an Abelian-like electric field $\tilde{f}_{4i}$. How good an approximation it is to the real and expected flux squeezing of $\sqrt{\sum_a(E^a_i)^2}$ depends strongly on the choice of $n^a$. 

An Abelian projection adopting a special gauge called Maximally Abelian gauge (MA)~\cite{Suzuki:1983cg,Kronfeld:1987ri,Kronfeld:1987vd} is found to give us interesting results~\cite{Suzuki:1992rw,Chernodub:1997ay,Suzuki:1998hc}  supporting 
importance  of the Abelian monopoles. In this case, the Abelian electric field 
can approximate very well the long-range behavior of the non-Abelian one, since off-diagonal components are suppressed. However such beautiful results are not seen in other general gauges. 

It is the purpose of  this note to show numerically
that the  dual Meissner effect is observed in a gauge-invariant way with the use of  a  gauge-invariant Abelian-like field strength and  a monopole-like quantity. We do not need  any 
Abelian projection nor any gauge-fixing.
Monte-Carlo simulations of quenched $SU(2)$ QCD are performed. It is found that the squeezing of the non-Abelian electric field $\sqrt{\sum_a (E^a_i)^2}$
occurs and the solenoidal current from the gauge-invariant  monopole-like quantity is responsible for the flux squeezing. The magnetic displacement current observed previously in Landau gauge~\cite{Suzuki:2004dw} is found to be negligible.      Preliminary results are obtained with respect to the vacuum type of the confinement phase. The $SU(2)$ QCD vacuum seems  near the border between the type 1 and the type 2 dual superconductors.
The present numerical results are not perfect, since the continuum limit, the infinite-volume limit and  the real $SU(3)$ case  are not studied yet. 
 Nevertheless the authors think the results 
obtained here are very interesting to general readers, since they show for the first time the flux squeezing of non-Abelian electric fields  is working in a gauge-invariant way due to the  dual Meissner effect   without performing any Abelian projection.

Let us define an  Abelian-like  field strength and a gauge-invariant  monopole-like quantity  in QCD.  The   field strength is written in terms of a unit-vector in color space which is constructed by a non-Abelian field strength itself:
\begin{eqnarray}
f_{\mu\nu}(x)&=&\sum_a n^a_{\mu\nu}(x)F^a_{\mu\nu}(x). \label{fmunu}
\end{eqnarray}
 $F^a_{\mu\nu}$ is a non-Abelian field strength. $n^a_{\mu\nu}$ is a unit vector in color space transforming as an adjoint representation in $SU(2)$~\cite{Chernodub:2000wk,Chernodub:2000bq,Chernodub:2000rg} 
 and is a symmetric tensor in space
:
\begin{eqnarray}
n^a_{\mu\nu}(x)&=&\epsilon_{\mu\nu}F^a_{\mu\nu}(x)/\sqrt{\sum_b (F^b_{\mu\nu}(x))^2}, \label{eq.SU2}
\end{eqnarray}
where $\epsilon_{\mu\nu}$ is an antisymmetric tensor with a sign
convention $\epsilon_{\mu<\nu}=1$. The opposite sign convention can be
adopted.  Note that Eq.(\ref{fmunu}) is just equal to the
gauge-invariant absolute value of the non-Abelian field strength itself
except for the sign. Hence it is not a simple Lorenz tensor. 
It is noted that 
an electric field component $E_i$ defined by $f_{4i}$ is $-\sqrt{\sum_a (E^a_i)^2}$ the squeezing of which is to be explained.
 
A gauge-invariant   monopole-like quantity is defined by 
\begin{eqnarray}
k_{\mu}(x)=\frac{1}{8\pi}\epsilon_{\mu\nu\alpha\beta}\partial_{\nu}
f_{\alpha\beta}(x), \label{monopole}
\end{eqnarray}
which is conserved but is not a simple Lorenz vector. Hereafter we call the monopole-like quantity simply as 'monopole'.
We get from Eq.(\ref{monopole})
\begin{eqnarray}
\vec{\nabla}\times\vec{E}+\partial_{4}\vec{B}&=&4\pi\vec{k}, \label{BI}\\
\vec{\nabla}\cdot\vec{B}&=&-4\pi k_4, \label{BI-2}  
\end{eqnarray}
where $B_i\equiv 1/2\sum_{jk}\epsilon_{ijk}f_{jk}$ and the vector notation is with respect to the three-dimensional space.
Note that the magnetic charge 
 defined in Eq.(\ref{BI-2}) does not satisfy the Dirac quantization condition with respect to  bare charges contrary to the usual case of a magnetic charge defined in terms of 
a 'tHooft field strength\footnote{ Even if we extend the definition Eq.(\ref{fmunu}) to a form like a 'tHooft field strength, the 'monopole' does not become topological, since $n^a_{\mu\nu}$ depends on $\mu$ and $\nu$. Almost the same numerical data were, however, obtained with this extended definition~\cite{Suzuki:2005ab}.}. 


\begin{figure}[b]
\includegraphics[height=5cm]{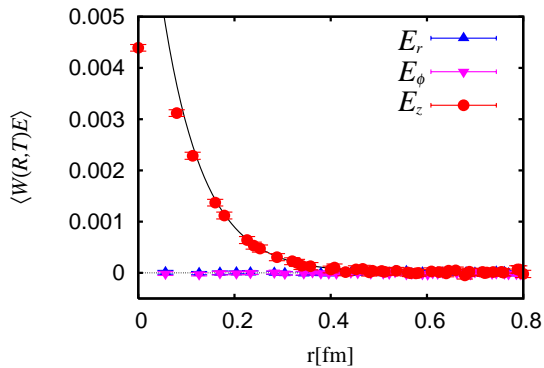}
\caption{\label{fig-1} $\vec{E}$ electric field  profiles. $r$ is a distance perpendicular to the $Q\bar{Q}$ axis and 
 $W(R\times T= 5\times 5)$ is used. 
The solid line denotes the best exponential fit.
}
\end{figure}

Now we go to a lattice QCD framework and perform numerical simulations in pure $SU(2)$ QCD. We adopt an improved Iwasaki gluonic action~\cite{Iwasaki:1985we}.
 Here we use  thermalized 2000 vacuum configurations at the 
 lattice distance $a=0.0792(2)$ fm. 
Simulation details are the same as in Ref.\cite{Suzuki:2004dw}.

A non-Abelian field strength $F_{\mu\nu}(s)$ is 
given by a $1\times 1$ plaquette variable defined by 
a path-ordered product of four non-Abelian link matrices on the lattice:
\begin{eqnarray*}
U_{\mu\nu}(s)= \exp\left(iF_{\mu\nu}(s)\right)= U^0_{\mu\nu}(s)+iU^a_{\mu\nu}(s)\sigma^a.
\end{eqnarray*}
The unit vector in color space is 
\begin{eqnarray}
n^a_{\mu\nu}(s) = \epsilon^{\mu\nu}U^a_{\mu\nu}(s)/\sqrt{1-(U_{\mu\nu}^0(s))^2}\label{unit-vector}
\end{eqnarray}
and the Abelian-like field strength is written similarly as 
in Eq.(\ref{fmunu}), i.e., 
$f_{\mu\nu}(s)=\sum_a n^a_{\mu\nu}(s)F^a_{\mu\nu}(s)$.

Let us try to measure, without any gauge-fixing, electric and magnetic flux distributions by evaluating correlations of Wilson loops and the Abelian-like field strengths located in the perpendicular direction to the Wilson-loop plane.
 
 
\begin{figure}
\includegraphics[height=5.cm]{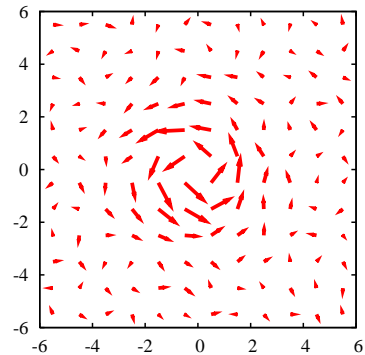}
\caption{\label{fig-2}Monopole current distributions.}
\end{figure}


We  define a gauge-invariant lattice 'monopole' in the same way as in Eq.(\ref{monopole}):
\begin{eqnarray}
k_{\mu}(s)&=&\frac{1}{8\pi}\epsilon_{\mu\nu\alpha\beta}\Delta_{\nu}
f_{\alpha\beta}(s+\hat{\mu}), \label{lattice-monopole}
\end{eqnarray}
which satisfies $\Delta'_{\mu}k_{\mu}(s)=0$.  $\Delta_{\mu}$ ($\Delta'_{\mu}$) is a lattice forward (backward) derivative.
  


\begin{figure}
\includegraphics[height=6cm,width=8.5cm]{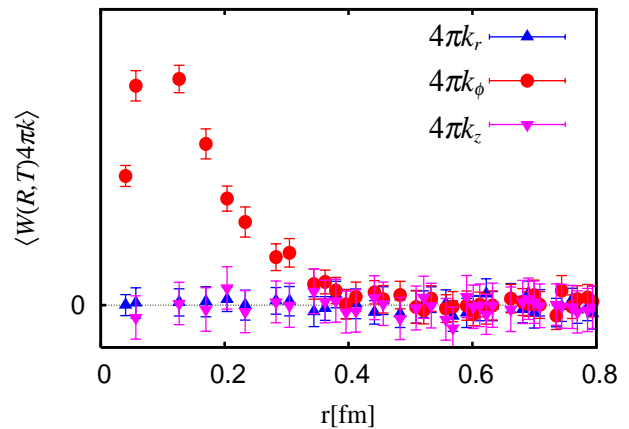}
\caption{\label{fig-4}Components of  monopoles around a static quark pair.}
\end{figure}

\begin{figure}
\includegraphics[height=6cm,width=8.5cm]{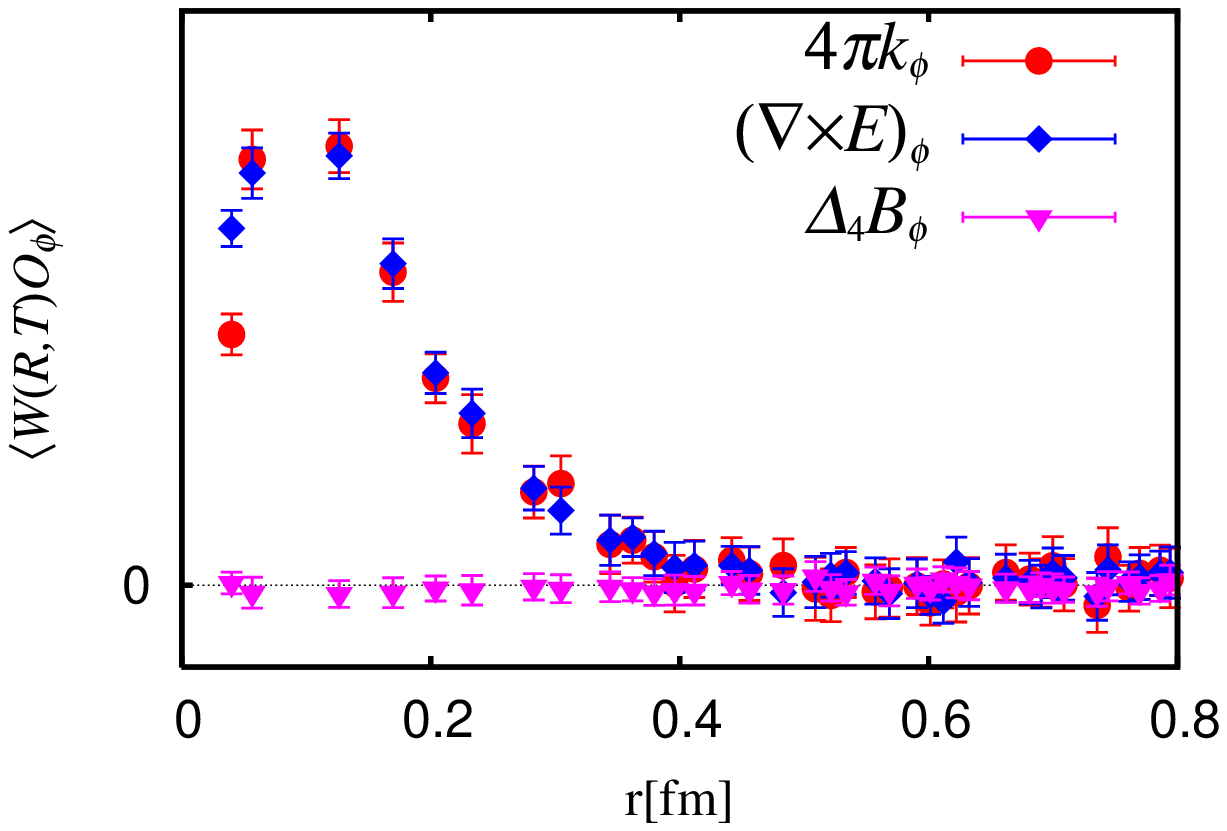}
\caption{\label{fig-5}The azimuthal component of the Abelian monopoles and the magnetic displacement current  around a static quark pair.}
\end{figure}

\begin{figure}
\includegraphics[height=5cm]{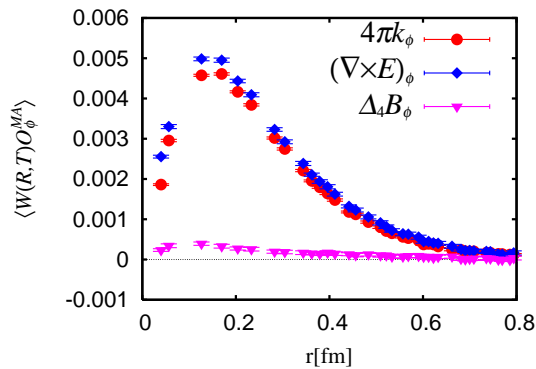}
\caption{\label{fig-6}The azimuthal component of the Abelian monopoles and the magnetic displacement current  around a static quark pair in the MA gauge.}
\end{figure}

First we show in Fig.\ref{fig-1}  electric  field profiles around a 
quark pair. Only the $z$-component of the electric field is non-vanishing and squeezed.
The profiles are studied mainly on a perpendicular plane at the midpoint between the 
two quarks. Note that electric fields perpendicular to the $Q\bar{Q}$ axis are found to be negligible. The solid line denotes the best exponential fit.


\begin{figure}[htb]
\includegraphics[height=4.7cm]{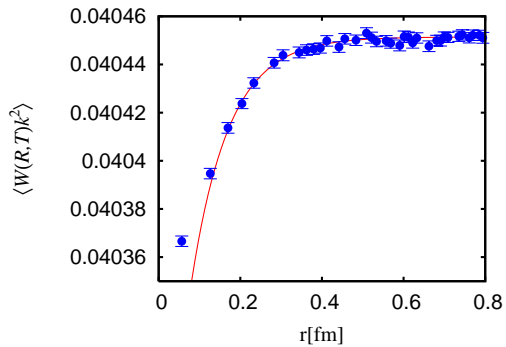}
\caption{\label{fig-7} The correlation between the Wilson loop and the
 squared monopole density. $W(R=5,T=5)$ is used. 
The solid line denotes the best exponential fit.
}
\end{figure}
\begin{figure}[htb]
\includegraphics[height=4.7cm]{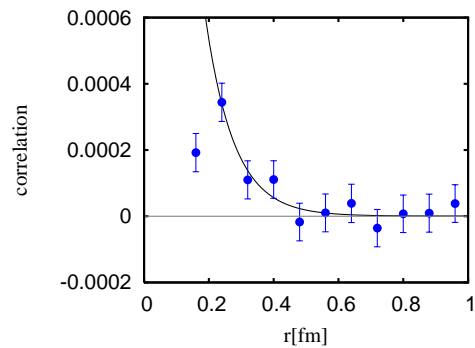}
\caption{\label{fig-8} Behaviors of $\Delta_r\Delta'_r E_z+
 (1/r)\Delta_{r}E_z$. 
The solid line denotes the best exponential fit.
}
\end{figure}

Now let us study the violation of the Bianchi identity with respect to the Abelian-like field strength Eq.(\ref{BI}). 
The Coulombic electric field coming from the static source is written in the lowest perturbation theory in terms of the gradient of a scalar potential.  Hence it does not contribute to the curl of the  electric field nor to the  magnetic field in the above  Bianchi identity  Eq.(\ref{BI}). 
The dual Meissner effect  says that the squeezing of the electric flux occurs due to cancellation of the Coulombic electric fields and those from  solenoidal magnetic currents. It is very interesting to see from Fig.\ref{fig-2} that in this gauge-invariant case, the gauge-invariant 'monopole' Eq.(\ref{lattice-monopole}) plays the role of the solenoidal current. This is qualitatively similar to  the monopole behaviors in the MA gauge~\cite{Singh:1993jj,Bali:1996dm,Koma:2003gq,Koma:2003hv}. 
 
 Let us see also the $r$ dependence of the 'monopole' distribution shown
 in  Fig.\ref{fig-4}. All $r$ and $z$ components of each term in
 Eq.(\ref{BI}) are almost vanishing  consistently with
 Fig.\ref{fig-2}. The magnetic displacement current $\Delta_4\vec{B}$
 are found to be negligible numerically as similarly as in the MA
 gauge~\cite{Singh:1993jj,Cea:1995zt,Bali:1996dm}. 
We have measured also correlations between  Wilson loops and  electric currents 
defined by $j_{\mu}=\Delta_{\nu}f_{\mu\nu}$.  They are found to be negligible. 

For the sake of comparison, we also discuss the MA gauge where 
$\sum_{s,\mu} Tr [ U_{\mu}(s)\sigma_3U_{\mu}^{\dagger}(s)\sigma_3]$
is maximized. In the MA gauge, an Abelian link variable $\theta_{\mu}^{MA}(s)$ is defined by a phase 
of the diagonal part of a non-Abelian link field after the gauge-fixing.
An Abelian field strength $\theta_{\mu\nu}^{MA}(s)$ in MA gauge is defined as
$\theta_{\mu\nu}^{MA}(s)\equiv\theta_{\mu}^{MA}(s)+\theta^{MA}_{\nu}(s+\hat{\mu})-\theta_{\mu}^{MA}(s+\hat{\nu})-\theta^{MA}_{\nu}(s).$ In this case, we use only the third component $W^3$ of the non-Abelian Wilson loop as a source.
We show in Fig.\ref{fig-5} and Fig.\ref{fig-6}
azimuthal components of  all three terms of Eq.(\ref{BI})
in this gauge-invariant case and in the MA-gauge case.  
It is interesting that the peak positions of gauge-invariant $k_{\phi}$
and $k_{\phi}^{MA}$ in the MA gauge look similar around $0.15$ fm, although the height and 
the shapes seem  different. Note that in the MA gauge, 
$W^3$ 
are used as a source of a quark and an anti-quark. 

Let us next try to fix the type of the vacuum of pure $SU(2)$ QCD.
The penetration length $\lambda$ is determined by making an exponential fit to the electric field flux for large $r$ regions.
 The best fitting curve is also plotted in Fig.~\ref{fig-1} from which we fix the penetration length $\lambda $
\footnote{We have fixed both lengths using a simple exponential function expected in the long-range regions. For details, see Ref.\cite{Chernodub:2005gz}.}.
Next  we derive the coherence length $\xi$. In a separate paper~\cite{Chernodub:2005gz},
we have shown that the coherence length can be fixed by a measurement of
the squared monopole density 
$k^2(s)\equiv \sum_{\mu=1}^4 k_\mu^2(s)$
 around the $Q\bar{Q}$ pair. The same situation is expected in this gauge-invariant case. 
The correlation between the Wilson loop and the squared 'monopole' density  is plotted in
Fig.~\ref{fig-7}. From the exponential fit, we may fix the coherence length. We get
$\lambda=0.085(5)$ fm and 
$\xi/\sqrt{2}=0.10(1)$ fm.
Although  the $R$ and $T$ dependences of both lengths are not studied
 yet, the value of the coherence length looks almost the same as that of
 the penetration length within the error bars.  Hence if the same
 situations will continue for larger $R$  in the confining string
 region, the type of the vacuum is fixed to be near the border between
 the type 1 and the type 2. 
This is consistent with the result of our previous
paper~\cite{Chernodub:2005gz} and the  results (see ~\cite{Haymaker:2005py} and references therein) obtained in the MA gauge and Landau gauge. 
 Also similar results were obtained  in SU(3) dual QCD in the continuum~\cite{Baker:1989qp}.
It should be stressed, however, that the present result is obtained in  the gauge-invariant framework. 

Next we study an equation similar to the dual London equation.
From Eq.(\ref{BI}), we get
\begin{eqnarray*}
4\pi\vec{\nabla}\times\vec{k}&=&\vec{\nabla}(\vec{\nabla}\cdot\vec{E})-\triangle\vec{E}+\vec{\nabla}\times(\partial_4\vec{B}). 
\end{eqnarray*}
Evaluating correlations between a Wilson loop and each term in the righthand side,
we find only $z$ components are relevant:
$4\pi(\vec{\nabla}\times\vec{k})_z \sim-\Delta_{r}\Delta'_{r}E_z-(1/r)\Delta_{r}E_z
$. Fig.\ref{fig-8} shows the data with the use of 4000 thermalized
configurations.  The exponential fit can be done for large $r\ge 3a$,
although the errors are large. The correlation length $0.09(2)$ fm is
compatible with the penetration length. We also see
$4\pi(\vec{\nabla}\times\vec{k})_z \sim m^2 E_z$
 with $m^{-1}=0.10(4)$ fm~\cite{Haymaker:2005py} which is also compatible with the penetration length.

Finally two comments are in order:
(1)  The Abelian-like field strength Eq.(\ref{fmunu}) is reduced to an Abelian one if off-diagonal components are negligible or non-Abelian $1\times 1$ plaquettes are well approximated by Abelian ones. The former 
occurs in the MA gauge, whereas the latter does in the maximally Abelian Wilson loop gauge~\cite{Shoji:1999gj} where almost the same fine results as in the MA gauge are observed. In these cases, we may adopt $n_{\mu\nu}\sim 1$, since only the diagonal components are dominant. Hence the present gauge-invariant results could explain why only restricted Abelian projection schemes like the MA gauge look nice among infinite possible candidates.\\
(2)  It is very interesting that the non-Abelian action is written by $f_{\mu\nu}^2$ which can be decomposed into 'monopole' and 'electric-current' parts with the help of the Hodge decomposition. Using the Michael action sum-rule~\cite{Michael:1986yi}, we can evaluate the 'monopole' contribution to the static potential. The work is in progress.
 
The numerical simulations of this work were done using RSCC computer clusters in 
RIKEN. The authors would like to thank RIKEN for their support of computer facilities. 
T.S. is supported by JSPS Grant-in-Aid for Scientific Research on Priority Areas 13135210 and (B) 15340073.

\bibliographystyle{../ref/h-physrev3}
\bibliography{../ref/confine,../ref/suzuki-mod,../ref/maxim-mod,../ref/fedor,../ref/bali}

\begin{thebibliography}{10}

\bibitem{Bali:1994de}
G.~S. Bali, K.~Schilling, and C.~Schlichter,
\newblock Phys. Rev. {\bf D51}, 5165 (1995), hep-lat/9409005.

\bibitem{tHooft:1975pu}
G.~'t~Hooft,
\newblock in {\em Proceedings of the EPS International}, edited by A.~Zichichi,
  p. 1225, 1976.

\bibitem{Mandelstam:1974pi}
S.~Mandelstam,
\newblock Phys. Rept. {\bf 23}, 245 (1976).

\bibitem{'tHooft:1974qc}
G.~'t~Hooft,
\newblock Nucl. Phys. {\bf B79}, 276 (1974).

\bibitem{Polyakov:1974ek}
A.~M. Polyakov,
\newblock JETP Lett. {\bf 20}, 194 (1974).

\bibitem{tHooft:1981ht}
G.~'t~Hooft,
\newblock Nucl. Phys. {\bf B190}, 455 (1981).

\bibitem{Suzuki:1983cg}
T.~Suzuki,
\newblock Prog. Theor. Phys. {\bf 69}, 1827 (1983).

\bibitem{Kronfeld:1987ri}
A.~S. Kronfeld, M.~L. Laursen, G.~Schierholz, and U.~J. Wiese,
\newblock Phys. Lett. {\bf B198}, 516 (1987).

\bibitem{Kronfeld:1987vd}
A.~S. Kronfeld, G.~Schierholz, and U.~J. Wiese,
\newblock Nucl. Phys. {\bf B293}, 461 (1987).

\bibitem{Suzuki:1992rw}
T.~Suzuki,
\newblock Nucl. Phys. Proc. Suppl. {\bf 30}, 176 (1993).

\bibitem{Chernodub:1997ay}
M.~N. Chernodub and M.~I. Polikarpov,
\newblock 
\newblock in {\em "Confinement, Duality and Nonperturbative Aspects of QCD"},
  edited by P.~van Baal, p. 387, Cambridge, 1997, Plenum Press.

\bibitem{Suzuki:1998hc}
T.~Suzuki,
\newblock Prog. Theor. Phys. Suppl. {\bf 131}, 633 (1998).

\bibitem{Suzuki:2004dw}
T.~Suzuki, K.~Ishiguro, Y.~Mori, and T.~Sekido,
\newblock Phys. Rev. Lett. {\bf 94}, 132001 (2005), 

\bibitem{Chernodub:2000wk}
M.~N. Chernodub, F.~V. Gubarev, M.~I. Polikarpov, and V.~I. Zakharov,
\newblock Nucl. Phys. {\bf B592}, 107 (2001), 

\bibitem{Chernodub:2000bq}
M.~N. Chernodub, F.~V. Gubarev, M.~I. Polikarpov, and V.~I. Zakharov,
\newblock Phys. Atom. Nucl. {\bf 64}, 561 (2001), 

\bibitem{Chernodub:2000rg}
M.~N. Chernodub, F.~V. Gubarev, M.~I. Polikarpov, and V.~I. Zakharov,
\newblock Nucl. Phys. {\bf B600}, 163 (2001), 

\bibitem{Iwasaki:1985we}
Y.~Iwasaki,
\newblock Nucl. Phys. {\bf B258}, 141 (1985).

\bibitem{Singh:1993jj}
V.~Singh, D.~A. Browne, and R.~W. Haymaker,
\newblock Phys. Lett. {\bf B306}, 115 (1993).

\bibitem{Bali:1996dm}
G.~S. Bali, V.~Bornyakov, M.~Muller-Preussker, and K.~Schilling,
\newblock Phys. Rev. {\bf D54}, 2863 (1996).

\bibitem{Koma:2003gq}
Y.~Koma, M.~Koma, E.-M. Ilgenfritz, T.~Suzuki, and M.~I. Polikarpov,
\newblock Phys. Rev. {\bf D68}, 094018 (2003), 

\bibitem{Koma:2003hv}
Y.~Koma, M.~Koma, E.-M. Ilgenfritz, and T.~Suzuki,
\newblock Phys. Rev. {\bf D68}, 114504 (2003), 

\bibitem{Cea:1995zt}
P.~Cea and L.~Cosmai,
\newblock Phys. Rev. {\bf D52}, 5152 (1995).

\bibitem{Chernodub:2005gz}
M.~N. Chernodub {\em et~al.},
\newblock Phys. Rev. {\bf D72}, 074505 (2005), 

\bibitem{Haymaker:2005py}
R.~W. Haymaker and T.~Matsuki,
\newblock (2005), hep-lat/0505019.

\bibitem{Baker:1989qp}
M.~Baker, J.~S. Ball, and F.~Zachariasen,
\newblock Phys. Rev. {\bf D41}, 2612 (1990).

\bibitem{Shoji:1999gj}
F.~Shoji, T.~Suzuki, H.~Kodama, and A.~Nakamura,
\newblock Phys. Lett. {\bf B476}, 199 (2000), 

\bibitem{Michael:1986yi}
C.~Michael,
\newblock Nucl. Phys. {\bf B280}, 13 (1987).

\bibitem{Suzuki:2005ab}
T.~Suzuki, K.~Ishiguro, Y.~Nakamura, and T.~Sekido,
\newblock Gauge invariant 'monopoles' and color confinement mechanism,
\newblock in {\em Proceedings of Lattice 2005 conference}, 2005.

\end{thebibliography}

\end{document}